\documentclass[10pt,twocolumn,letterpaper]{article}

\usepackage{cvpr}

\usepackage{amsmath}
\usepackage{amsfonts}
\usepackage{amssymb}
\usepackage{graphicx}
\usepackage{siunitx}
\usepackage[euler]{textgreek}
\DeclareMathOperator*{\argmin}{arg\,min}
\newcommand{\norm}[1]{\left\lVert#1\right\rVert}
\newcommand\Paragraph[1]{\vspace{2mm}
\textbf{#1}   }
\usepackage{cite}
\usepackage{url}
\usepackage{lipsum}
\usepackage{amsmath}

\usepackage{array, makecell} 
\usepackage{blindtext}
\usepackage[utf8]{inputenc}
\usepackage[table]{xcolor}
\usepackage[flushleft]{threeparttable} 
\usepackage{subcaption}
\usepackage{soul}
\usepackage{tabularx,ragged2e,booktabs,caption}
\usepackage{adjustbox}
\usepackage[usestackEOL]{stackengine}
\cvprfinalcopy

\title{Recent Advances in Imaging Around Corners}
\author{Tomohiro Maeda, Guy Satat, Tristan Swedish, Lagnojita Sinha, Ramesh Raskar\\
MIT Media Lab}
\date{}

\begin{document}

\twocolumn[{%
\renewcommand\twocolumn[1][]{#1}%
\maketitle

\begin{center}
 \includegraphics[width=1.0\textwidth]{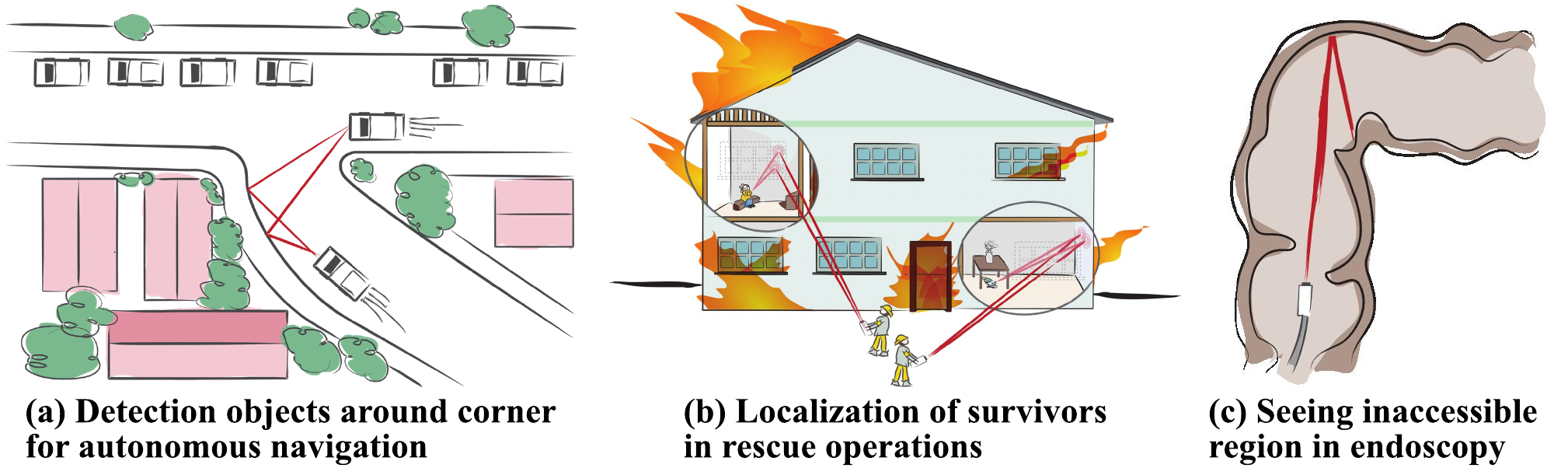}
 \captionof{figure}{\textbf{Applications of non-line-of-sight (NLOS) imaging. (Material from:` \cite{CORNARweb}, adapted with permission from the authors.')} (a) Detection of approaching vehicles around corners results in safer autonomous navigation. (b) Localizing survivors in the hidden areas makes search-and-rescue operations more efficient and safer. (c) Imaging the inaccessible regions in endoscopy enables better diagnostics.}
 \label{fig:applications}
\end{center}
}]

\begin{abstract}
Seeing around corners, also known as non-line-of-sight (NLOS) imaging is a computational method to resolve or recover objects hidden around corners. Recent advances in imaging around corners have gained significant interest. This paper reviews different types of existing NLOS imaging techniques and discusses the challenges that need to be addressed, especially for their applications outside of a constrained laboratory environment. Our goal is to introduce this topic to broader research communities as well as provide insights that would lead to further developments in this research area.
\end{abstract}

\section{Introduction}

\begin{figure*}
        \centering
        \includegraphics[width=1.0\linewidth]{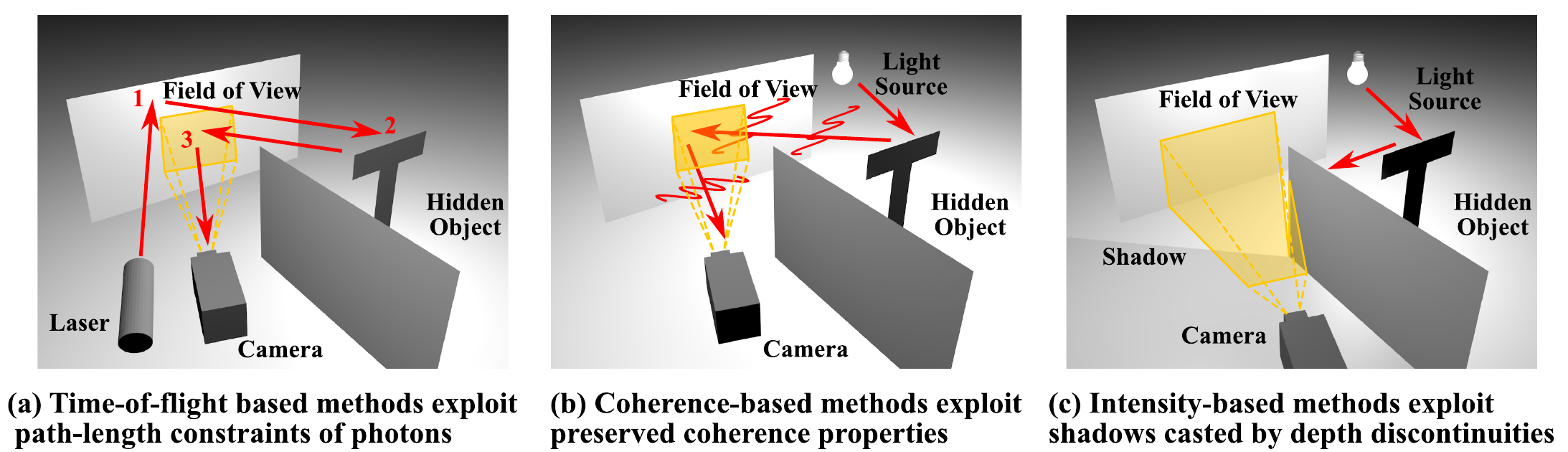}
        \caption{\textbf{Layouts of typical NLOS imaging setups with different sensing principles.} (a) ToF-based techniques use three-bounce photons that give distance constraints. (b) Coherence-based techniques use speckle patterns or spatial coherence that are preserved during scattering. (c) Intensity-based methods mainly exploit occlusions that cast shadows.} 
    \label{fig:CORNAR_setup}
    \vspace{-2mm}
\end{figure*}

\begin{figure*}
        \centering
        \includegraphics[width=1.0\linewidth]{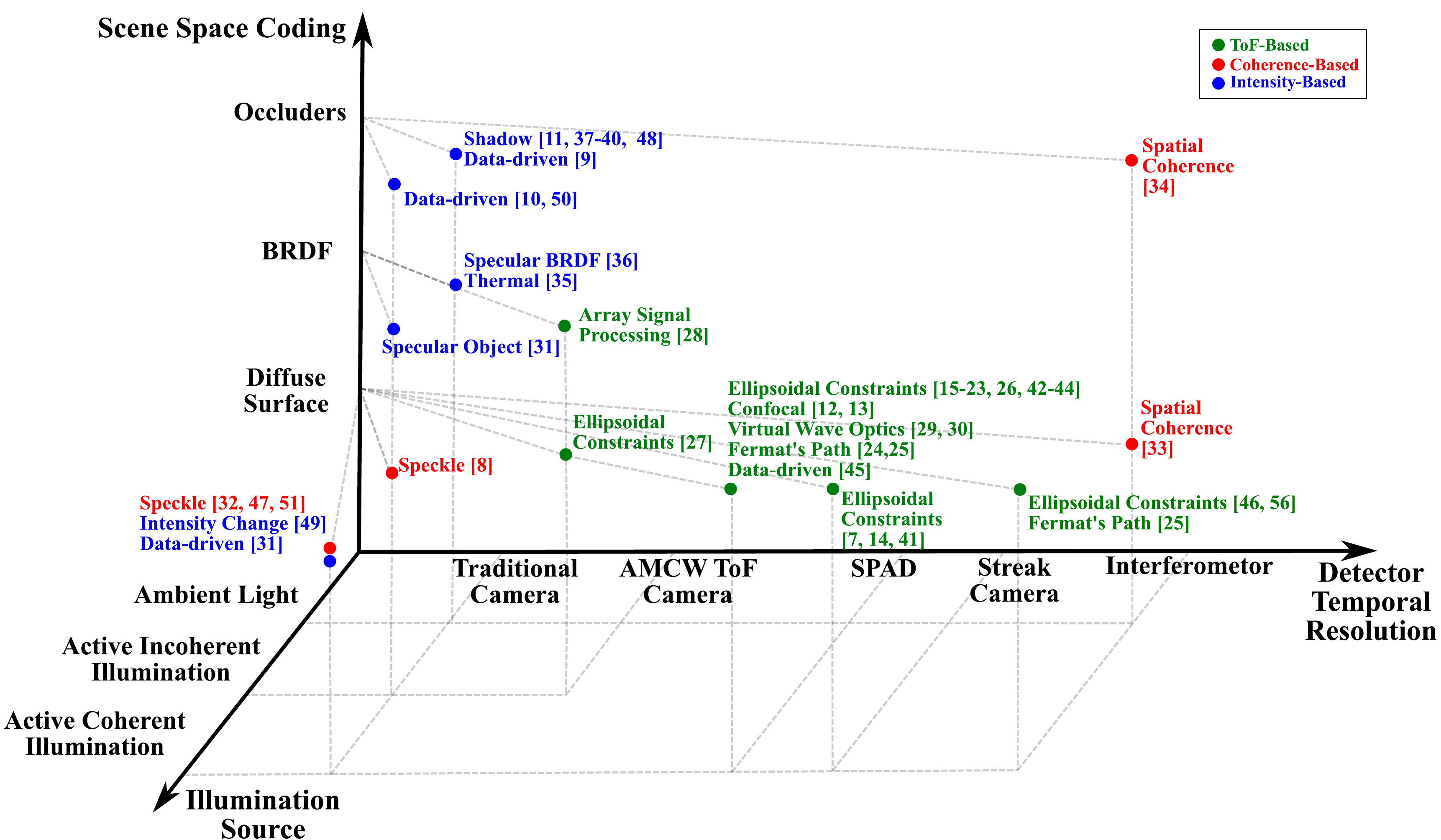}
        \caption{\textbf{Landscape of NLOS imaging works on detector, illumination source, and scene space coding.}} 
    \label{fig:CORNAR_Summary}
    \vspace{-2mm}
\end{figure*}

Recent advances in computational imaging techniques made it possible to image around corners, which is known as non-line-of-sight (NLOS) imaging. The ability to see around corners would be beneficial in various applications. For example, the detection of objects around a corner enables autonomous vehicles to avoid collisions. Detection and localization of people without the need to go into dangerous environments make rescue operations safer and more efficient. NLOS imaging can also be used for medical applications such as endoscopy, where the region of interest is hard for the sensors to access directly (Fig.~\ref{fig:applications}). 

Fig.~\ref{fig:CORNAR_setup} shows typical scene setups for NLOS imaging techniques. While there are many techniques using the electromagnetic (EM) spectrum or acoustic waves to image around corners and through walls~\cite{Adib2013Wifi,Adib:2015,Lindell:2019:Acoustic}, we focus on works that use light (electromagnetic wave in the visible and infrared spectrum).  Fig.~\ref{fig:CORNAR_Summary} shows an overview of the current state of NLOS imaging.

NLOS imaging was first proposed by Raskar and Davis~\cite{Raskar5DT}, and demonstrated by Kirmani et al.~\cite{Kirmani09} for recovery of hidden planar patches. Velten et al.~\cite{Velten12} showed the first full reconstruction of a 3D object that was hidden around a corner. 
These works showed that ToF measurements of photons returning from the hidden scene with three bounces (Fig.~\ref{fig:CORNAR_setup} (a)) contain sufficient information to recover the occluded scene. Following these works, different sensing modalities were used for NLOS imaging. For example, Katz et al.~\cite{Katz14} first demonstrated the use of a speckle pattern to reconstruct 2D images. Tancik et al.~\cite{Tancik2018cosi,Tancik2018FlashPF} and Bouman et al.~\cite{Bouman17} demonstrated reconstruction and tracking of the hidden object with a traditional RGB camera.

The main challenge in NLOS imaging is that the photons reflected from the hidden object are scattered at the line-of-sight surfaces (e.g., wall and floor). Computational imaging techniques model light transport to recover the information of the hidden scene. However, their applications in practice are still limited. This paper reviews the recent advances in NLOS imaging and discusses the challenges towards real-world applications. We note that other types of imaging tasks, such as seeing through scattering media, are often also referred to as NLOS imaging, but in this paper, we use the term ``NLOS imaging'' to refer solely to seeing around corners.

%%%%%%%%%%%%%%%%%%%%%%%%%%%%%%%%%%%%%%%%%%%%%%%%%%%%%%%%%%%
%Begin Table
%%%%%%%%%%%%%%%%%%%%%%%%%%%%%%%%%%%%%%%%%%%%%%%%%%%%%%%%%%%
\begin{table*}
    \centering
    \resizebox{\textwidth}{!}{
    \begin{tabular}{c|c|c|c|}\cline{2-4}
         & \Centerstack{ \\ \textbf{ToF} \\} & \textbf{Coherence} & \textbf{Intensity} \\ 
         \hline
         \multicolumn{1}{|c|}{\Centerstack{\textbf{3D} \\\textbf{Reconstruction}}}& \cellcolor[RGB]{175,250,175}\Centerstack{\textmu m-cm Resolution \\\cite{Velten12,OToole:2018:ConfocalNLOS,Heide:2019:OcclusionNLOS,Gupta12,Buttafava15,Arellano17,Laurenzis14,Manna2018ErrorBA,Laurenzis:15,Jin:18,Pediredla17,Iseringhausen2018NonLineofSightRU,Adithya2019:SNLOS,Tsai17,Xin:19,tsai2019beyond,Heide14DiffuseMirror,Kadambi16,liu2019phasor_nlos,Lindell:2019:Wave}} & \cellcolor[RGB]{250,175,175} None &\cellcolor[RGB]{250,250,175} \Centerstack{Planar/Specular Object~\cite{chen_2019_nlos}} \\ 
         \hline
          \multicolumn{1}{|c|}{\makecell{\textbf{2D} \\\textbf{Reconstruction}}}& \cellcolor[RGB]{175,250,175} Included in 3D & \cellcolor[RGB]{175,250,175}\Centerstack{Diffraction Limited\\ Resolution~\cite{Katz14,Viswanath:18}\\ cm Resolution\cite{Batarseh18,Beckus2018MultimodalNP}} &\cellcolor[RGB]{250,250,175} \Centerstack{Coarse Resolution~\cite{chen_2019_nlos}, Thermal~\cite{ICCP19_Maeda} \\ Occlusion Dependent~\cite{Tancik2018cosi,Tancik2018FlashPF,Sasaki18,Baradad2018InferringLF,Thrampoulidis2018ExploitingOI,Saunders2019Periscopy,Yedidia_2019_CVPR}} \\ 
         \hline
          \multicolumn{1}{|c|}{\makecell{\textbf{Localization/} \\\textbf{Tracking}}}& \cellcolor[RGB]{175,250,175} \Centerstack{cm Precision\\\cite{Pandharkar11,Gariepy:16,Chan17FastTracking,Chan2017NonlineofsightTO, Caramazza2018NeuralNI,Boger-Lambard18}}  & \cellcolor[RGB]{250,175,175} \cellcolor[RGB]{250,250,175} \Centerstack{1D Distance Recovery\cite{Batarseh18}\\3D Tracking~\cite{Smith_2018_CVPR}}&\cellcolor[RGB]{250,250,175} \Centerstack{Occlusion Dependent~\cite{Bouman17,Tancik2018FlashPF, Seidel2019} \\ 6D Tracking~\cite{Klein-2016-Tracking}, 3D localization~\cite{ICCP19_Maeda,Chandran2019}}\\ 
         \hline
         \multicolumn{1}{|c|}{\Centerstack{\textbf{Classification}}}& \cellcolor[RGB]{175,250,175} \Centerstack{Human \\ Identification~\cite{Caramazza2018NeuralNI}}& \cellcolor[RGB]{175,250,175}\Centerstack{MNIST, \\ Human Pose Classification~\cite{Lei_2019_CVPR}}&\cellcolor[RGB]{175,250,175} \Centerstack{Object Type Classification~\cite{Tancik2018FlashPF,Chandran2019},\\ Human Pose Detection~\cite{ICCP19_Maeda}}  \\ 
         \hline
    \end{tabular}
    }
     \caption{\textbf{Overview of existing techniques to see around corners for different tasks.} The color of the table cell indicates the quality of the methods relative to other NLOS imaging techniques. Green and yellow cells indicate that NLOS imaging tasks have been shown in practical and limited recovery complexities. Red cells indicate that the tasks have not been demonstrated.} 
    \label{tab:comp}
\end{table*}

\subsection{What is NLOS Imaging?}
Fig.~\ref{fig:CORNAR_setup} illustrates different strategies to see around corners. Mathematically, NLOS imaging can be formulated as the following forward model:
\begin{equation}\label{eq:function}
    \mathbf{y} = f(\mathbf{x}) + \epsilon,
\end{equation}
where, $\mathbf{x}$ represents the hidden scene parameters, such as albedo, motion or class of the hidden object, $\mathbf{y}$ is the measurement, $f(\cdot)$ is the mapping of the hidden scene to the measurement--which depends on the choice of illumination, sensor, and the geometry of  the corner--and $\epsilon$ denotes the measurement noise.

The goal of NLOS imaging is to design sensing schemes such that $f(\cdot)$ can be inverted, and to build algorithms such that $\mathbf{x}$ is robustly and efficiently recovered given $\mathbf{y}$. Recently, data-driven approaches showed that inversion of Eq.~\ref{eq:function} can be learned even when $f(\cdot)$ is not directly available~\cite{Tancik2018FlashPF}.

\Paragraph{Full Scene Reconstruction:} The objective of NLOS imaging is to see around a corner as if there is a line-of-sight directly into the hidden scene. Reconstruction tasks include the recovery of the 3D, 2D, or 1D shape of the hidden object or light fields. In reconstruction, $\mathbf{x}$ is a direct representation of the hidden scene, such as a voxelized probability map, surface, or light transport in the hidden scene.

\Paragraph{Direct Scene Inference:} The reconstruction procedure often requires the long acquisition and time-consuming computations. However, full scene reconstruction might not even be necessary in many scenarios--for example, detecting the presence of a moving person is sufficient for rescue operations. We define ``inference'' as an estimation of these latent parameters (e.g., location, class, or motion) without direct reconstruction of the hidden scene.  In this case, the unknown parameter $\mathbf{x}$ can be location, type, or movement of the hidden object.   Because inference does not require full reconstruction, inference around corners can be potentially performed with less computation and measurement than full scene reconstruction.

\subsection{Overview of Sensing Schemes for NLOS Imaging}
We classify the existing techniques into three categories based on their sensing modalities: Time-of-flight, coherence, and intensity. Table 1 provides an overview of the current state of each method regarding NLOS imaging tasks. Section 2--4 provides the review of NLOS imaging techniques that exploit different sensing modalities, and Section 5 presents the challenges of NLOS imaging for each category.

\Paragraph{Time-of-Flight-Based Approach (Section 2):}
 As the photons travel from the source, the first bounce occurs on an adjoining wall or directly on the floor near the corner. The second bounce is on the hidden object around the corner. The third bounce is again on a wall or a floor in the line-of-sight region and back towards the measurement system, as shown in Fig.~\ref{fig:CORNAR_setup} (a). Time-resolved imaging techniques utilize the photons associated with the third bounce, which constrains possible locations of the hidden object. Section~\ref{sec:ToFSection} reviews the reconstruction and inference algorithms for time-resolved sensing based methods.

\Paragraph{Coherence-Based Approach (Section 3):}
While the information on the hidden geometry is largely lost by the diffuse reflection of photons on the relay wall, some coherence properties of light are preserved. Coherence-based methods exploit the fact that coherence preserves information about the occluded scene. Section~\ref{sec:coherence_based} reviews NLOS imaging techniques that utilize speckle and spatial coherence of light.

\Paragraph{Intensity-Based Approach (Section 4):}
Occlusions from the corner or occlusions within the hidden scene can provide rich information on the hidden scene. Using occlusions, it is possible to recover the hidden scene with a typical intensity (RGB) camera, such as a smartphone camera (Fig.~\ref{fig:CORNAR_setup} (c)). While many intensity-based approaches rely on occlusions, some works demonstrate NLOS imaging by exploiting surface reflectance of the relay wall or the object. Section~\ref{sec:intensity_based} reviews the intensity-based approach.

\subsection{Cameras to Image Around Corners}\label{sec:ToFHardware}
Different hardware has been used to see around corners. We briefly introduce such hardware to describe what sensors and illumination sources are used for NLOS imaging.

\Paragraph{Streak Camera + Pulsed Laser:} 
A streak camera acquires temporal information by mapping the arrival time of photons in 1D space with sweep electrodes, resulting in a 2D measurement of space and time. Streak cameras offer the best time resolution down to sub-picosecond or even 100 femtoseconds (sub-mm --0.03 mm distance resolution), among other types of ToF cameras. The disadvantages of streak cameras are the high cost (typically more than \$100k), low photon efficiency, high noise level, and the need for line scanning or additional optics~\cite{Liang14}. This results in a potentially longer acquisition time for sufficient signal-to-noise ratio (SNR). The first works on NLOS imaging used a Hamamatsu C5680 for the experiments with Kerr lens mode-locked Ti:sapphire laser (50 fs pulse width).

\Paragraph{SPAD + Pulsed Laser:} 
Single-photon avalanche diode (SPAD) can detect the arrival of a single photon with a time jitter of 20--100 ps (6--30 mm distance resolution). Unlike streak cameras, it is possible to have 2D SPAD arrays for capturing 3D measurements without the need for line scanning. While the temporal resolution of a SPAD is poorer than streak cameras, the photon efficiency and SNR are better. Commonly used single-pixel SPAD detectors and time-correlated single-photon counting devices are from Micro Photon Devices and Hydraharp from PicoQuant. These cost around \$5k and \$20-35k, respectively.

\Paragraph{AMCW ToF Camera + Modulated Light Source:} An amplitude modulated continuous wave (AMCW) ToF camera compares the phase shift of emitted and received modulated light through three or more correlation measurements~\cite{Bttgen05, Buttgen08}. The distance resolution of AMCW ToF cameras is limited by modulation frequency and the SNR. While AMCW ToF cameras can achieve few-mm resolution~\cite{Yang15}, they are susceptible to strong ambient light due to the need for a longer exposure time than that for pulse-based ToF cameras. AMCW ToF cameras have a much lower cost (ranging \$400--\$2000) than SPADs and streak cameras and are used in commercial products, including the Microsoft Kinect and Photonic Mixer Device (PMD).

\Paragraph{Traditional Camera:} In this paper, we use ``traditional camera'' to refer to the sensors such as charge-coupled device (CCD) and complementary metal-oxide-semiconductor (CMOS) arrays, which measure irradiance images without ToF information. Traditional cameras can be used for both intensity and speckle measurement modalities. The combination of a diffuser and traditional camera enables the measurement of cross-correlation of temporal coherence to provide passive ToF measurement~\cite{Boger-Lambard18}. Traditional cameras are the most ubiquitous and affordable sensor discussed in this paper but do not acquire direct measurements of time-of-flight information.

\Paragraph{Interferometer:} Interference between multiple lightwaves provides depth information in \textmu m scale, which can be used for ToF-based NLOS imaging for microscopic scenes. For example, an optical coherence tomography system with temporally and spatially incoherent LED showed 10 \textmu m resolution to demonstrate NLOS reconstruction of an object as small as a coin~\cite{Xin:19}. Heterodyne interferometry, which utilizes interference of light with different wavelengths, demonstrated NLOS imaging at 70 \textmu m precision~\cite{Willomitzer:18}. A dual-phase Sagnac interferometer captures complex-valued spatial coherence functions as a change of measured intensity and is used for spatial coherence-based methods. Such an interferometer is not commercially available, and we refer readers to \cite{RezvaniNaraghi:17} for the design and details of a dual-phase Sagnac interferometer. An interferometer is a sensitive instrument and often requires careful alignment and isolation from mechanical vibrations. 

\section{Time-of-Flight-Based NLOS Imaging}\label{sec:ToFSection}
Among many numbers of works in NLOS imaging, ToF-based techniques are the most popular due to their ability to resolve the path length of the three-bounce photons that carry the information of the hidden scene. The ToF measurement of three-bounce photons can also be used for estimating the bidirectional reflectance distribution function (BRDF) of materials~\cite{Naik:2011}. While this review paper only considers imaging around corners, ToF measurements are also useful for seeing through scattering media~\cite{Satat:15,Kumar07,Raviv:14,Satat:16,Satat:18}, analyzing light transport~\cite{Velten2013FemtophotographyCA,Kadambi:2013,Bhandari:14, Kadambi:14, Wu2014,Gariepy2015SinglephotonSL,Kadambi:16Interferometry, OToole:2017:SPAD}, and novel imaging systems~\cite{Satat:16Compressive, Heshmat:16, Heshmat:18}.

\begin{figure}
        \centering
        \includegraphics[width=0.98\linewidth]{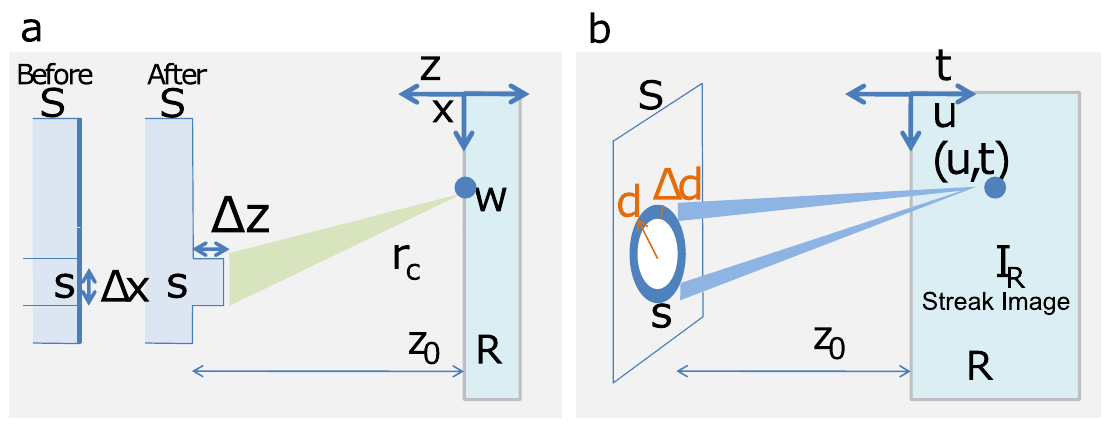}
        \caption{\textbf{Need for time-resolved information (Material from Velten et al., “Recovering three-dimensional shape around a corner  using  ultrafast  time-of-flight  imaging,” published  2012,   Nature  Communications~\cite{Velten12} with permission of the authors)}. a) Intensity change due to displacement of a patch is below traditional camera's sensitivity. (b) Temporal information resolves the displacement of a patch with a practical sensitivity range.} 
    \label{fig:time_vs_intensity}
    \vspace{-2mm}
\end{figure}

\Paragraph{The Benefit of ToF Measurement}
As discussed in the supplement material of Velten et al.~\cite{Velten12}, the change of intensity due to the displacement of a patch illustrated in Fig.~\ref{fig:time_vs_intensity} (a) is proportional to $(\Delta x)^2\Delta z / z^3$, where $\Delta_x ,\Delta_z, z$ are the size, displacement and depth of a patch respectively. In contrast, the intensity change of a time-resolved measurement as illustrated in Fig.~\ref{fig:time_vs_intensity} (b) is proportional to $3 \Delta x / (2\pi z)$. Drawing upon an example from ~\cite{Velten12}, the fractional change of the intensity for $\Delta_x$ = 5 mm, $\Delta_z$ = 5mm, z = 20 cm is 0.00003 for the traditional camera and 0.01 for the ToF camera. This example shows that intensity change is below the sensitivity of the traditional camera. Furthermore, the SNR is small because the number of three-bounce photons is small. For this reason, the existing demonstration of intensity-based NLOS imaging relies on occlusions or more specular BRDF of the wall. ToF measurement can resolve the change of the measurement to recover the 3D geometry of the hidden object.

\subsection{Reconstruction Algorithms}\label{sec:ToFReconstruct}

\begin{figure}
        \centering
        \includegraphics[width=0.8\linewidth]{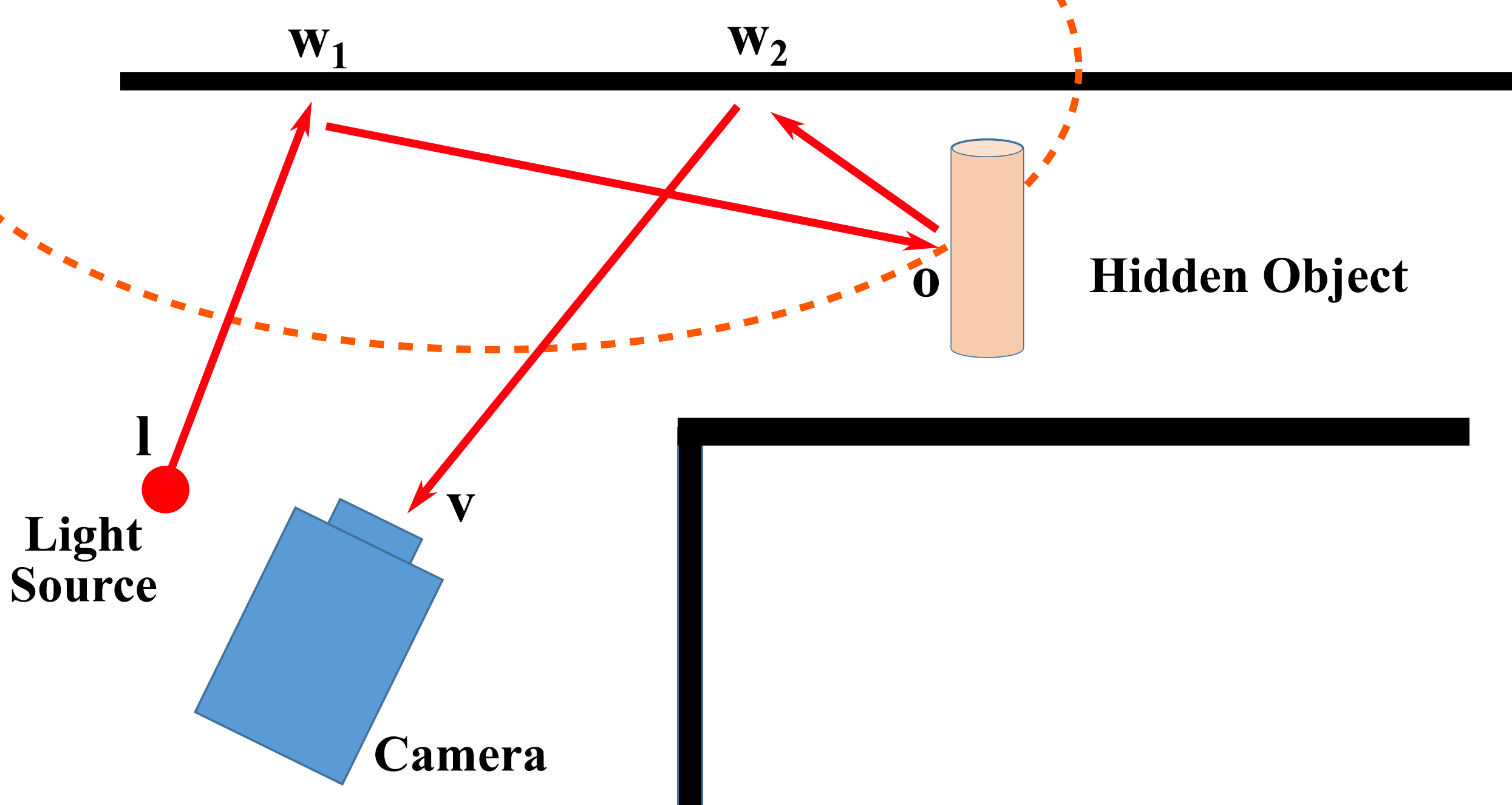}
        \caption{\textbf{Time-of-flight information gives constraints on the potential location of the object around corners.} When the path length of a photon is known, the location of the hidden object is constrained to the ellipsoid.} 
    \label{fig:NLOS_ToF}
    \vspace{-2mm}
\end{figure}

ToF measurements provide ellipsoidal constraints on the possible object locations in the hidden scene, as illustrated in Fig.~\ref{fig:NLOS_ToF}. Let $w_1$ and $w_2$ denote points of the wall where a photon undergoes the first and third bounces, and $o$ denote a point of the object where a photon undergoes the second bounce. Moreover, let $l$ and $v$ denote locations of the light source and the camera. Then the hidden object $o$ must be on an ellipsoid that satisfies 
\begin{equation}
    \lvert w_1 - o  \rvert + \lvert w_2 - o  \rvert = ct - \lvert w_1 - l  \rvert - \lvert w_2 - v  \rvert,
\label{eq:ellipsoid}
\end{equation}
where $c$ and $t$ are the speed of light and time of travel respectively.  

\begin{figure}
        \centering
        \includegraphics[width=0.85\linewidth]{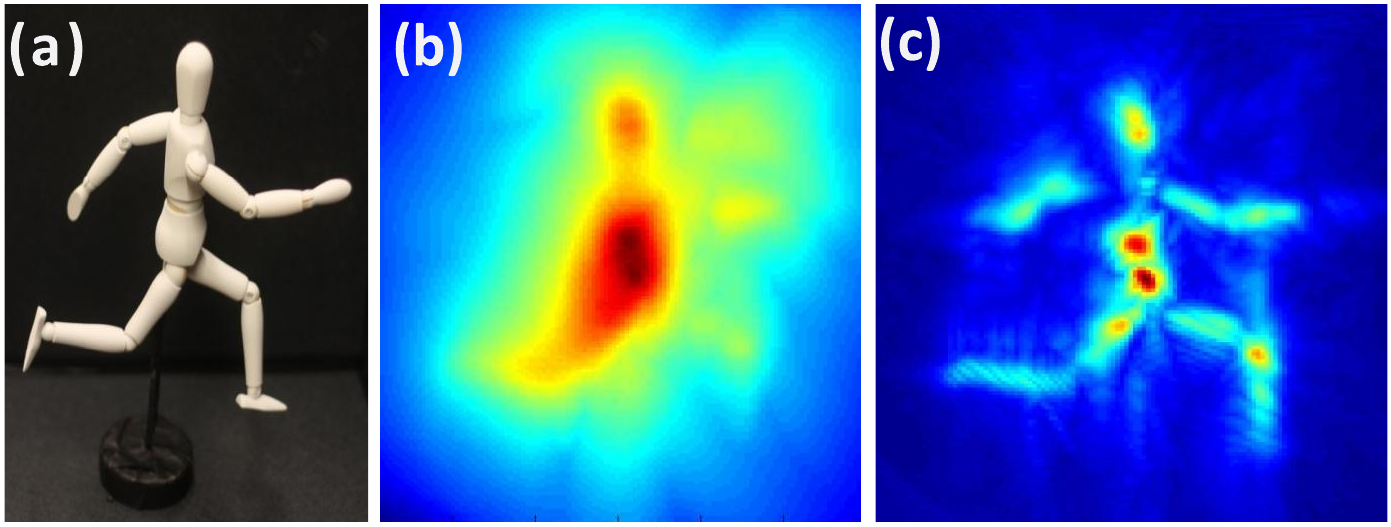}
        \caption{\textbf{Visualization of reconstruction via back-projection (Material from: Velten et al., ``Recovering three-dimensional shape around a corner using ultrafast time-of-flight imaging,'' published 2012, Nature Communications~\cite{Velten12} adapted with permission of SNCSC).} (a) The hidden object around the corner. (b) Probability map generated by a back-projection algorithm. (c) The contour of the object can be sharpened by applying a sharpening filter.}
    \label{fig:back-projection}
    \vspace{-2mm}
\end{figure}

\begin{figure*}
        \centering
        \includegraphics[width=0.7\linewidth]{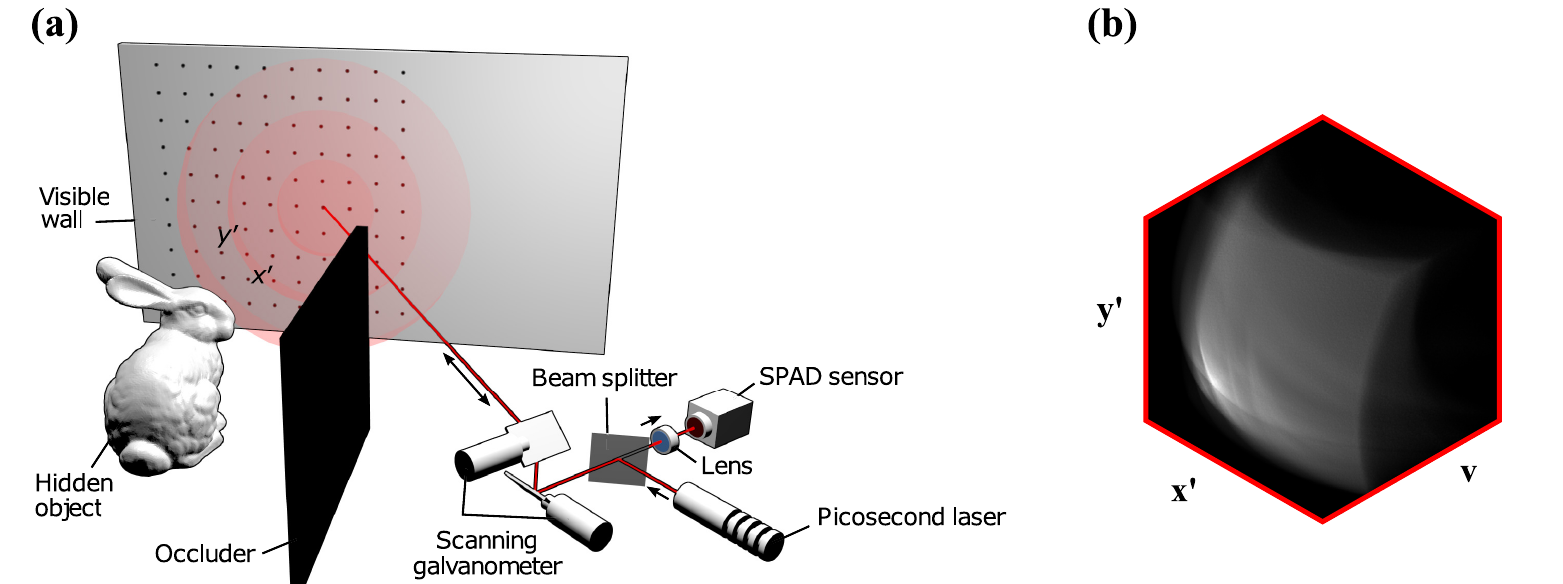}
        \caption{\textbf{Layout of Confocal imaging setup (Material from: O'Toole et al., ``Confocal non-line-of-sight imaging based on the light-cone transform,'' published 2018, Nature~\cite{OToole:2018:ConfocalNLOS}, adapted with permission of SNCSC).} (a) The SPAD pixel and laser sees the same point on the wall. Due to the confocal optical design and change of variables, the measurement can be written as a convolution of the hidden object and a 3D kernel. (b) 3D kernel can be used to perform deconvolution to recover the hidden object. } 
    \label{fig:confocal}
    \vspace{-2mm}
\end{figure*}

Most of the existing techniques consider discretized voxels to recover hidden geometry. This physics-based forward model maps the hidden scene to the measurement. Many works express the forward model with the ellipsoidal constraint as a linear inverse problem, which can be solved by back-projection or optimization.
\begin{equation}
    \mathbf{y} = \mathbf{A}\mathbf{x} + \epsilon,
\label{eq:setup}
\end{equation}
where $\mathbf{x} \in \mathbb{R}^n$ and $\mathbf{y} \in \mathbb{R}^m$ and $\epsilon \in \mathbb{R}^m$ denote the vectorized target voxels, ToF measurements, and noise respectively. $A \in \mathbb{R}^{m\times n}$ represents the transient light transport including the ellipsoidal constraints described by Eq.~\ref{eq:ellipsoid}, intensity fall-off due to the surface reflectance, and the distance between the object and the wall. The theory of light transport of multi-bounce photons is studied by Seitz et al.~\cite{Seitz:2005} for non-time-resolved imaging, and by Raskar et al.~\cite{Raskar5DT} for time-resolved imaging.

In NLOS imaging, Eq.~\ref{eq:setup} is often ill-posed, so filtering or regularization of the reconstruction is often necessary. Moreover, $\mathbf{A}$ becomes large as $x$ and $y$ are 5D measurements (3D measurement $\times$ 2D scanning) and 3D voxels in general. For example, if we have $n_x \times n_y \times n_t$ 3D measurement for scanning illumination with $n_s$ points, $i$th column of $A_i$ has $ n = n_x n_y n_t n_s$ elements, which maps the voxel $x_i$ to the measurement. If we take $64\times 64\times \times 64$ 3D measurement with $64$ laser illumination spots for $64\times 64\times 64$ voxel reconstruction, $A$ is a $64^4$ by $64^3$ matrix. ToF-based methods mainly focus on efficient and robust algorithms to recover $\mathbf{x}$.

Recently, a forward model beyond a linear model has been studied. For example, the shape of the surface can be recovered using fewer photons by modeling photons that travel specific paths called Fermat paths~\cite{Xin:19}. Phasor-field virtual wave optics enables the modeling of full light transport, including photons that bounce more than three times, in the hidden scene~\cite{liu2019phasor_nlos}.

\Paragraph{Back-Projection:}
A naive way to solve Eq.~\ref{eq:setup} is to consider each voxel (an element in $\mathbf{x}$), and compute the heat map of an object occupying the voxel given the measurement $\mathbf{y}$ and light transport model $\mathbf{A}$. This back-projection method was exploited in the first demonstration of the 3D reconstruction of hidden objects~\cite{Velten12}, and other NLOS imaging works for reconstruction~\cite{Gupta12, Buttafava15, Laurenzis14, Manna2018ErrorBA, Laurenzis:15, Jin:18}. Back-projection usually produces blurry reconstruction, as shown in Fig.~\ref{fig:back-projection} (b) because of the ill-posed nature of $\mathbf{A}$. Sharpening filters and thresholding are used to improve the reconstruction quality. Instead of considering each voxel, a probability map of $\mathbf{x}$ can be recovered by considering the intersections of ellipsoidal constraints to perform efficient back-projection that is up to three orders of magnitude faster than naive back-projection~\cite{Arellano17}. Back-projection can be implemented optically, by illuminating and scanning along ellipsoids on the relay wall. This enables focusing the measurement to a single voxel in the hidden scene~\cite{Adithya2019:SNLOS}.

\Paragraph{Optimization:} 
Priors on the hidden object $\mathbf{x}$ can be incorporated to the inversion of Eq.~\ref{eq:setup} by formulating the inverse problem as minimization of a least square error with a regularizer~\cite{Gupta12,Heide14DiffuseMirror,Heide:2019:OcclusionNLOS,Kadambi16,Pediredla17}:
\begin{equation}
    \hat{\mathbf{x}} = \argmin_{x}\norm{\mathbf{y}-\mathbf{Ax}}^2 + \Gamma(\mathbf{x}),
\end{equation}
where $\Gamma(\mathbf{x})$ encourages $\mathbf{x}$ to follow priors of the hidden scene. Iterative algorithms such as CoSaMP~\cite{Needell10}, FISTA~\cite{Beck09:FISTA}, and ADMM~\cite{Boyd11} can be used to solve this optimization problem based on the priors. Enforcing priors generally results in better reconstruction quality as compared to back-projection.

While computation and memory complexity can be enormous, the optimization formulation gives flexibility in more accurate reconstruction. For example, $\mathbf{A}$ can be factorized to consider partial occlusion and surface normal of the hidden object to recover the visibility and surface normal as well as the albedo of the hidden object~\cite{Heide:2019:OcclusionNLOS}.

\Paragraph{Confocal Imaging:}
In confocal imaging, the relay wall is raster-scanned as the detector collects photons from the same point as the illuminated spot~\cite{OToole:2018:ConfocalNLOS}. This makes $w_1=w_2$, and the ellipsoidal constraints become spherical constraints, which makes the forward operation a 3D convolution (Fig.~\ref{fig:confocal}). Confocal setup relaxes the need to solve back-projection or optimization problems, and instead, a simple deconvolution solves the reconstruction problem. 3D deconvolution makes confocal imaging both memory- and computationally-efficient. Confocal imaging makes the inverse problem simple but suffers from the first-bounce photons as the detection and illumination points are the same. This issue can be mitigated by introducing a slight misalignment of the detector and illumination or time gating of SPAD. However, SNR is still limited because the single-bounce light is much stronger than the three-bounce light.

The above three methods formulate NLOS imaging as a linear inverse problem. Other problem formulations can be constructed to solve NLOS imaging from different perspectives.

\Paragraph{Wave-Based reconstruction:}
Reza et al.~\cite{reza2018physical} showed that the intensity waves from modulated light in the NLOS setting can be modeled as a propagation of a wave (phasor field). The hidden scene's impulse response can be recorded with a pulsed laser and ultrafast detector such as SPAD. The modulated light pulse can be virtually synthesized over time, using the recorded impulse response. Constructive interference of the synthesized wave appears at the hidden object. Hence, virtual propagation of a pulsed modulation on the impulse response of the hidden scene results in reconstruction~\cite{liu2019phasor_nlos}. Virtual wave optics approach to NLOS imaging models the full light transport, including more than three-bounce photons in the hidden scene. Lindel et al.~\cite{Lindell:2019:Wave} modeled the light transport of the confocal imaging system as wave propagation, where the measurement is one boundary condition. Acquiring other boundary conditions of the wave propagation with frequency-wavenumber (f-k) migration algorithm results in efficient and robust reconstruction of the hidden scene containing objects with various surface reflectance.

\begin{figure*}
        \centering
        \includegraphics[width=0.8\linewidth]{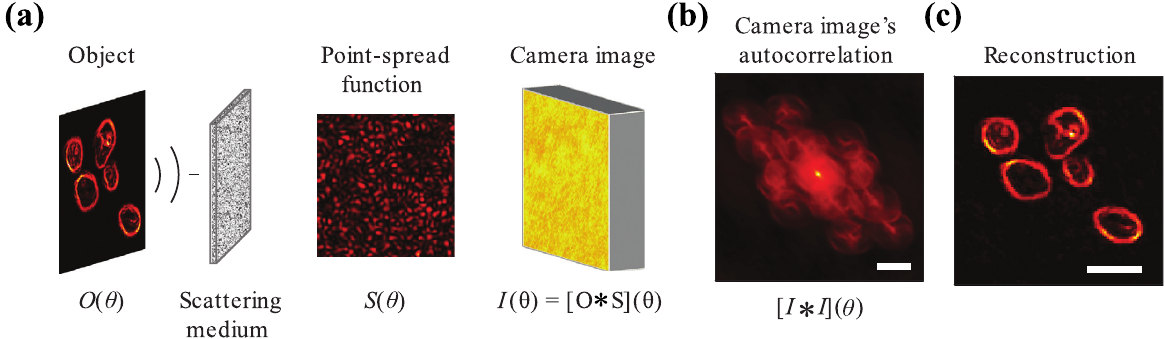}
       \caption{\textbf{The principle of speckle-based imaging technique to image after diffuse scattering (Material from: Katz et al., ``Non-invasive single-shot imaging through scattering layers and around corners via speckle correlations,'' published 2014, Nature Photonics~\cite{Katz14}, adapted with permission of SNCSC).} (a) Camera captures speckle pattern after scattering on a diffuse wall. (b) The angular auto-correlation of the captured image is the same as the angular auto-correlation of the original image. (c) Phase-retrieval algorithms recover the image before the scattering on the wall.} 
    \label{fig:speckle}
    \vspace{-2mm}
\end{figure*}

\Paragraph{Inverse Rendering:}
A renderer can be used to model the physics-based forward model instead of analytical forward operations, as written in Eq.~\ref{eq:setup}. This ``synthesis-by-analysis'' approach, also known as inverse rendering, changes the scene parameters such that the rendered and experimental measurements match. Inverse rendering provides more flexible reconstructions than the voxel-based reconstruction discussed above. For example, more detailed reconstruction is possible by representing the hidden object in mesh, and non-Lambertian surface reflection can be incorporated~\cite{Iseringhausen2018NonLineofSightRU}. However, accurate rendering can be time-consuming, especially because each iteration of the optimization requires computationally expensive rendering. The long run-time problem can be solved by a differential renderer that efficiently computes gradients with respect to the hidden scene parameters~\cite{tsai2019beyond}.

\Paragraph{Shape Recovery:}
Methods discussed above use full ToF measurement for reconstruction. However, the surface can be recovered without using all the multi-bounce photons. Tsai et al. showed that the first-returning photons provide the length of the shortest path to the hidden object, which can be used to reconstruct the boundary and surface normal of the hidden object~\cite{Tsai17}. Later, discontinuities of ToF measurement are shown to follow specific paths (Fermat paths) that give rich information about the boundary of the hidden geometry~\cite{Xin:19}. Because this approach does not rely on intensity information, it is robust to BRDF variations of the object around corners.

\subsection{Inference Algorithms for Localization, Tracking, and Classification}\label{sec:ToFInference}
We have reviewed algorithms to reconstruct objects around the corners. Often, reconstruction of the hidden object might not be the end goal. Instead, the inference of the properties of the hidden object, such as its location or class is sufficient. While reconstruction suffices such tasks, direct inference without reconstruction can be made more efficiently with a smaller number of measurements.

\Paragraph{Back-Projection:}
The back-projection algorithm that we discussed in the reconstruction section can be used for a point localization. Instead of considering small voxels to recover the details of the object, an object can be treated as the single voxel to recover the location of the object~\cite{Gariepy:16, Chan17FastTracking}. Because the goal is not to recover the 3D shape, localization can be performed with much fewer measurements and less computation than reconstruction at a larger scale ~\cite{Chan17FastTracking}. While reconstruction of the cluttered scene is challenging, tracking and size estimation of a moving object is demonstrated by Pandharka et al.~\cite{Pandharkar11}.

\Paragraph{Deep learning:} Data-driven algorithms have become a powerful tool for pattern recognition and found promising applications in computer vision. Though data-driven approaches for full reconstruction require further theoretical development~\cite{Gregor2010LearningFA,Chen18LISTA}, deep learning is useful for inference of unknown parameters such as object class and location. The objective of deep learning approach is to learn $x = f^{-1}(y)$, where $f(\cdot)$ could be hard to model explicitly. For example, neural networks with a SPAD array demonstrated the point localization and identification of a person around the corner~\cite{Caramazza2018NeuralNI}. While the imaging setup is different from a corner, Satat et al.~\cite{Satat:17} demonstrated calibration-free NLOS classification of objects behind a diffuser, where the transmissive scattering at the diffuser is similar to the reflective scattering at the relay wall.

\section{Coherence-based NLOS Imaging}\label{sec:coherence_based}
The challenges of NLOS imaging come from the scattering of photons on the diffuse relay wall. However, some coherent properties of light are preserved after scattering. Coherence-based methods exploit speckle patterns or spatial coherence to see around corners.

\subsection{Speckle-Based NLOS Imaging:}
The speckle pattern is an intensity fluctuation generated by the interference of the coherent light waves. Although a speckle pattern may seem random, the observed pattern encodes information about the hidden scene.

\Paragraph{Reconstruction}
The angular correlation of the object intensity pattern is preserved in the observed speckle pattern after the scattering on the relay wall~\cite{Isaac90}. This is known as memory effect~\cite{Feng88, Freund88}:
\begin{equation}
    \mathbf{y} * \mathbf{y}  = \mathbf{x} * \mathbf{x},
\end{equation}
where $\mathbf{y}$ and $\mathbf{x}$ are the observed speckle pattern and object intensity pattern, respectively. $*$ denotes convolution. Katz et al.~\cite{Katz14} found that the memory effect can be applied to a spatially incoherent light source such as fluorescent bulb, and demonstrated single-shot imaging through scattering, and around corners. Reconstruction of the hidden object can be performed with phase-retrieval algorithms~\cite{Shechtman2014PhaseRW, Jaganathan2015PhaseRA}. Speckle pattern can also be generated with active coherent illumination when the speckle pattern cannot be observed in passive sensing~\cite{Viswanath:18}. While this approach achieves diffraction-limited resolution, its field of view is limited because of the memory effect(order of a few degrees of the angular field of view~\cite{Katz14}). Because scattering of the diffuse wall has a similar nature as the scattering of a diffuser, speckle-based reconstruction can be used to see through scattering media as well~\cite{Bertolotti2012NoninvasiveIT}.

\Paragraph{Inference for Tracking}
When coherent light illuminates objects around a corner, the light scattered from the objects creates a speckle pattern on the relay wall. When the object moves, the speckle pattern moves as well. The motion of the object can be tracked by computing the cross-correlation of two images taken at different times~\cite{Smith:2017}. Smith et al.~\cite{Smith_2018_CVPR} showed that this principle applies to NLOS imaging, and demonstrated motion tracking of the multiple hidden reflectors using the active coherent illumination. The speckle-based tracking has precision less than 10 \textmu m, but is currently limited to microscopic motions because of the small field of view of the memory effect~\cite{Smith_2018_CVPR}. The data-driven approach demonstrated MNIST and pose-estimation classification from speckle patterns~\cite{Lei_2019_CVPR}.

\subsection{Spatial-Coherence-Based NLOS Imaging:}
Spatial coherence refers to the correlation of the phase of a light wave at different observation points (while temporal coherence refers to the correlation at different observation time). Spatial coherence can be used to reconstruct the scene in the camera's field of view~\cite{Beckus:18}. Because the spatial coherence of light is preserved through the scattering at the diffuse wall, such reconstruction techniques can be applied to NLOS imaging~\cite{Batarseh18,Beckus2018MultimodalNP}.
Measurement of spatial coherence requires a unique imaging system such as the Dual-Phase Sagnac Interferometer (DuPSaI)~\cite{RezvaniNaraghi:17}. 

\begin{figure*}
        \centering
        \includegraphics[width=0.95\linewidth]{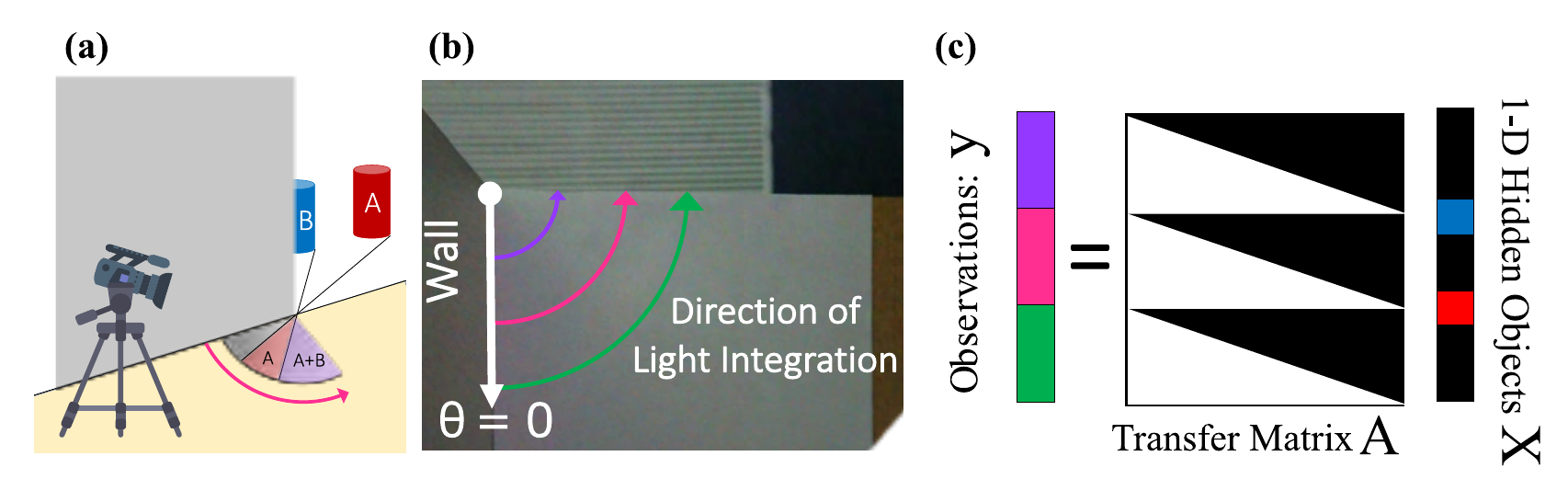}
        \caption{\textbf{The layout of NLOS imaging exploiting the scene occlusion (Material from: Bouman et al., ``Turning Corners into Cameras: Principles and Methods,'' published 2017~\cite{Bouman17} adapted with permission from IEEE. © 2017 IEEE).} (a) The corner occlusion casts shadows that depend on the location of the hidden objects. (b) Intensity camera sees the intensity over the angle from the relay wall. (c) Transfer (measurement) matrix $\mathbf{A}$ is constructed based on the occlusion.} 
    \label{fig:occlusion}
    \vspace{-2mm}
\end{figure*}
\begin{figure*}
        \centering
        \includegraphics[width=0.85\linewidth]{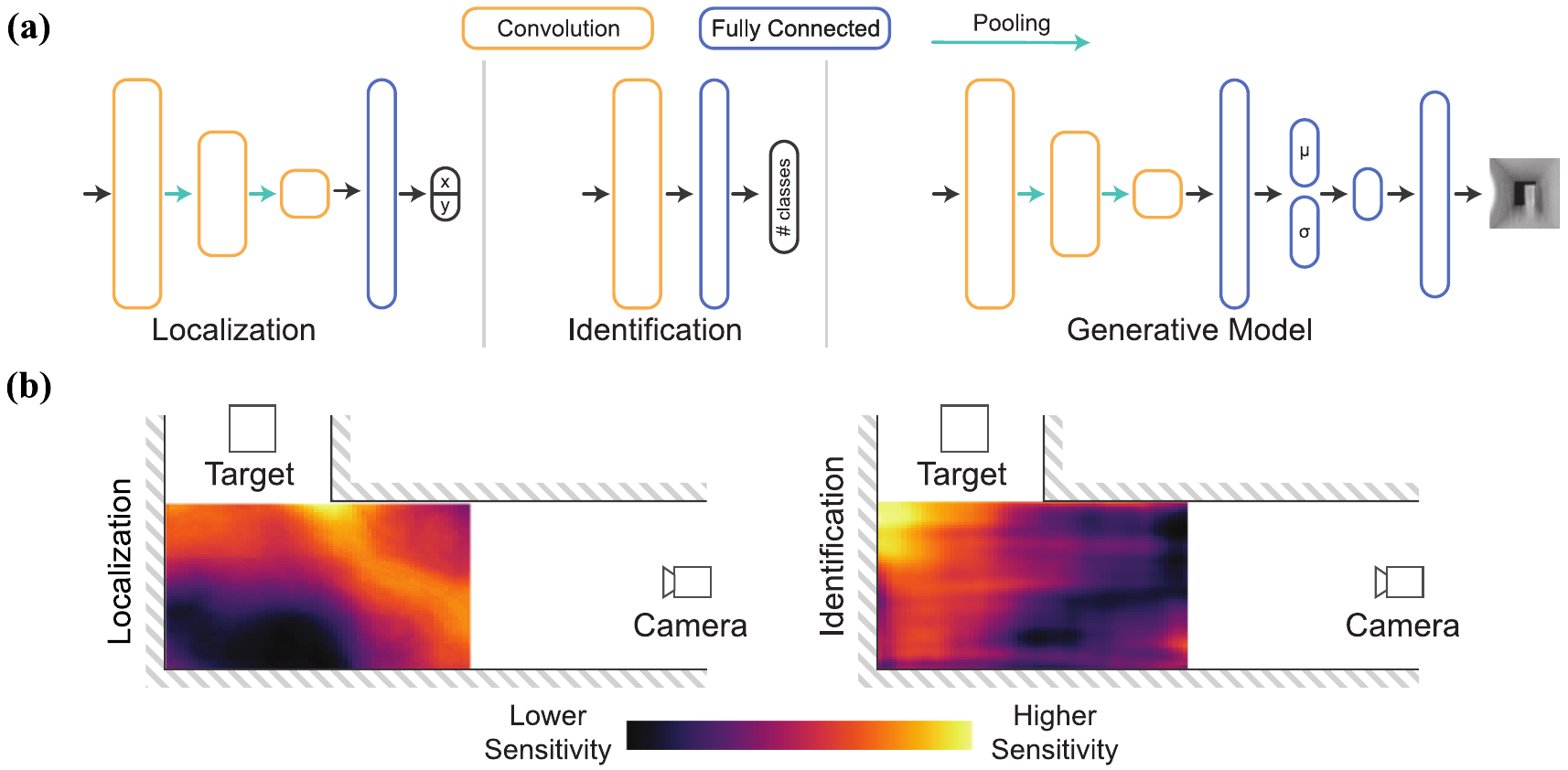}
        \caption{\textbf{Neural networks can locate, classify and reconstruct the hidden scene (Material from: Tancik et al., ``Flash Photography for Data-Driven Hidden Scene Recovery,'' published 2018~\cite{Tancik2018FlashPF} adapted with permission from the authors).} (a) Neural network architectures for different tasks. (b) Heatmaps of ``important'' region of the scene for localization and identification tasks. Neural networks learn to exploit the occlusion as proposed by \cite{Bouman17} (left). It also learns to use measurements that are previously not realized to be useful for object identification tasks (right).} 
    \label{fig:data_driven}
    \vspace{-2mm}
\end{figure*}

\paragraph{Reconstruction}
The propagation of coherence in free space can be approximated in a closed-form. This lets the spatial coherence measurement $\mathbf{y}$ be expressed as a system of linear equations as written in Eq.~\ref{eq:setup}, where $\mathbf{A}$ illustrates the propagation of the spatial coherence function through space and scattering. This inverse problem can be solved as minimization of least square error and regularizers that incorporate priors such as sparsity and small total variation. Beckus et al.~\cite{Beckus2018MultimodalNP} showed that coherence measurement and intensity measurement could be incorporated into a single optimization problem for reconstruction.

\Paragraph{Inference for Localization}
Using spatial coherence, the distance of the incoherent light source from the detector can be estimated from the phase of the spatial coherence function. While only 1D localization was demonstrated~\cite{Batarseh18}, multiple measurements of the spatial coherence functions can be used to localize the light source by triangulation.

\section{Intensity-Based NLOS Imaging}\label{sec:intensity_based}
Velten et al.~\cite{Velten12} first studied the use of intensity measurement for NLOS imaging and showed that a traditional camera requires high sensitivity for diffuse surfaces. We illustrate this in Fig.~\ref{fig:time_vs_intensity}, and refer readers to the supplemental material of \cite{Velten12} for further details. To overcome this challenge, most of the existing intensity-based NLOS imaging works exploit occlusions in the scene. Before the development of NLOS imaging, occlusions have been used for computational imaging for a long time. For example, light transport from occlusions are used to synthesize images from a different point of view~\cite{Sen:2005:DP}, spatially varying BRDF to capture incident light~\cite{Alldrin:2006:PLP}, recover 4D light field for refocusing~\cite{Veeraraghavan:2007}, and create an anti-pinhole camera to see outside the field-of-view of the camera~\cite{Torralba12:accidentalpinhole}. Tancik et al.~\cite{Tancik2018cosi, Tancik2018FlashPF} and Bouman et al.~\cite{Bouman17} first addressed the NLOS imaging problem directly. Bouman et al. showed 1D tracking, and Tancik et al. introduced the idea of using intensity-based reconstruction for NLOS imaging. We also review another class of technique that exploits non-Lambertian surface reflection.

\subsection{Exploiting Occlusions}
Spatially varying occlusions make it feasible to use intensity measurement since the variability of intensity due to occlusions increases the rank of the underlying ray transport matrix. Passive intensity-based NLOS imaging with inexpensive cameras has shown a real-time demonstration of seeing around corners. However, the separation of the ambient light and the signal light from the object is necessary, which often requires moving hidden objects or background subtraction.

%%%%%%%%%%%%%%%%%%%%%%%%%%%%%%%%%%%%%%%%%%%%%%%
% Begin Table
%%%%%%%%%%%%%%%%%%%%%%%%%%%%%%%%%%%%%%%%%%%%%%%
\begin{table*}\label{tab:comparingmethods}
\setlength{\arrayrulewidth}{.3 mm}
\setlength{\tabcolsep}{5pt}
\renewcommand{\arraystretch}{3}
\newcolumntype{s}{p{1cm}}
\arrayrulecolor[RGB]{0,0,0}
\fontsize{8}{8}\selectfont
\begin{center}
\begin{threeparttable}
\begin{tabular}{ c|c|c|c|c|c| }
\cline{2-6}
 & \Centerstack{\textbf{Illumination}} &\Centerstack{\textbf{Sensor}} &\Centerstack{\textbf{Cost}}&\Centerstack{\textbf{Sensitivity to}\\\textbf{Ambient Light}}&\Centerstack{\textbf{Need of Priors}\\ \textbf{on Geometry}}\\
\hline
\multicolumn{1}{|c|}{\Centerstack{\textbf{Pulsed} \\ \textbf{ToF}}} &  \Centerstack{Pulsed \\ Laser} & \Centerstack{Streak Camera \\ SPAD} & \cellcolor[RGB]{250,175,175}High  & \cellcolor[RGB]{175,250,175} Robust &  \cellcolor[RGB]{175,250,175} Not Required \\
\hline
\multicolumn{1}{|c|}{\Centerstack{\textbf{AMCW} \\ \textbf{ToF}}} &  \Centerstack{Modulated \\ Laser, LED} & \Centerstack{Correlation \\ Camera} &  \cellcolor[RGB]{250,250,175}Medium & \cellcolor[RGB]{250,250,175} \Centerstack{Sensitive to\\Strong Light} &\cellcolor[RGB]{175,250,175} Not Required \\
\hline
\multicolumn{1}{|c|}{\Centerstack{\textbf{Passive}\\\textbf{Coherence}}} & None  & \Centerstack{Traditional \\ Camera} & \cellcolor[RGB]{175,250,175} Low &\cellcolor[RGB]{250,175,175} Sensitive  & \cellcolor[RGB]{250,175,175}\Centerstack{Scene Geometry}\\
\hline
\multicolumn{1}{|c|}{\Centerstack{\textbf{Passive Spatial}\\\textbf{Coherence}}} & None  & \Centerstack{Dual Phase Sagnac \\ Interferometer }& \cellcolor[RGB]{250,250,175} Medium &\cellcolor[RGB]{250,175,175} Sensitive  & \cellcolor[RGB]{250,175,175}\Centerstack{Scene Geometry}\\
\hline
\multicolumn{1}{|c|}{\Centerstack{\textbf{Active}\\\textbf{Coherence}}} & \Centerstack{Coherent Source} & \Centerstack{Traditional \\ Camera}& \cellcolor[RGB]{175,250,175} Low & \cellcolor[RGB]{250,175,175} Sensitive &\cellcolor[RGB]{250,175,175}\Centerstack{Scene Geometry}\\
\hline
\multicolumn{1}{|c|}{\Centerstack{\textbf{Passive}\\\textbf{Intensity}}} & \Centerstack{None} & \Centerstack{Traditional \\ Camera}& \cellcolor[RGB]{175,250,175} Low & \cellcolor[RGB]{250,175,175} Sensitive &\cellcolor[RGB]{250,175,175}\Centerstack{Scene Geometry \\ Occlusion Geometry}\\
\hline
\multicolumn{1}{|c|}{\makecell{\textbf{Active}\\\textbf{Intensity}}} & \Centerstack{Flashlight \\ Laser} & \Centerstack{Traditional\\Camera}& \cellcolor[RGB]{175,250,175} Low & \cellcolor[RGB]{250,175,175} Sensitive & \cellcolor[RGB]{250,175,175}\Centerstack{Scene Geometry}\\
\hline
\end{tabular}
\end{threeparttable}
\caption{\textbf{Properties of techniques to see around corner for different sensing schemes.} This table summarizes fundamental properties of seeing around corners techniques.} 
\end{center}
\vspace{-3mm}
\end{table*}
%%%%%%%%%%%%%%%%%%%%%%%%%%%%%%%%%%%%%%%%%%%%%%%
% End Table
%%%%%%%%%%%%%%%%%%%%%%%%%%%%%%%%%%%%%%%%%%%%%%%

\Paragraph{Localization:}
Bouman et al.~\cite{Bouman17} demonstrated practical passive tracking of hidden objects using an occlusion from a wall as illustrated in Fig.~\ref{fig:occlusion}(a), and Tankic et al. showed that neural networks learn to exploit occlusions~\cite{Tancik2018FlashPF} as shown in Fig.~\ref{fig:data_driven} (b)    . The occlusion in the scene can be considered to be a specific type of aperture. As shown in Fig.~\ref{fig:occlusion} (c), the problem can be formulated as a linear inverse problem similar to Eq.~\ref{eq:setup}, where $\mathbf{A}$ represents light transport with occlusions. With priors on the floor albedo, it is possible to estimate the 1D angular location the hidden object from a single measurement by estimating the ambient illumination~\cite{Seidel2019}. 

\Paragraph{Reconstruction:}
More complex occlusion geometries enable reconstruction of the hidden scene, as $\mathbf{A}$ becomes well-posed~\cite{Thrampoulidis2018ExploitingOI,Saunders2019Periscopy}. A 4D light field can be recovered from an occlusion-based inverse problem framework~\cite{Baradad2018InferringLF}. While the above reconstruction techniques assume known occlusions, the unknown occlusions can be estimated by exploiting the sparse motion of the hidden object~\cite{Yedidia_2019_CVPR}. Deep learning-based approaches showed its ability for reconstruction, tracking, and object classification to a specific scene setup, while its generalizability is yet to be explored~\cite{Tancik2018FlashPF, Tancik2018cosi, Chandran2019}. 

\subsection{Exploiting Surface Reflectance}
Another class of intensity-based NLOS imaging technique exploits the bidirectional reflectance distribution function (BRDF) of the relay wall to reconstruct the hidden scene. Specular BRDF makes the inverse problem less ill-posed. When such reflectance function of the wall is known, the light field from the hidden scene can be reconstructed~\cite{Sasaki18}.  However, current demonstrations are limited to scenes without ambient light, and the wall has some specular surface reflections~\cite{Sasaki18}. Chen et al.~\cite{chen_2019_nlos} demonstrated the reconstruction of specular planar objects with active illumination. The data-driven approach showed that it is also possible to reconstruct diffuse objects with active illumination~\cite{chen_2019_nlos}. Tracking of the object using active illumination and intensity measurement can be performed by matching the simulation with the experimental measurements (inverse rendering)~\cite{Klein-2016-Tracking}. Thermal imaging benefits from the specular BRDF of long-wave IR, which was exploited for passive localization and near real-time pose detection around corners~\cite{ICCP19_Maeda}.

\section{Challenges and Future Directions}~\label{sec:challenges}
We reviewed major techniques for NLOS imaging in the previous sections. Here, we introduce common challenges to real-time and robust applications to inspire further research.

\subsection{Properties of Different Sensing Modalities}
Table 1 shows the current landscape of NLOS techniques, and Table 2 summarizes the properties of different sensing modalities. In this section, we discuss what challenges need to be solved to complete Table 1.

\Paragraph{ToF:} 
Since ToF-based NLOS imaging was first demonstrated, there have been significant improvements in reconstruction algorithms that became orders of magnitude faster than the originally proposed algorithm. The challenges towards practical ToF-based NLOS imaging are:
\begin{itemize}
  \item Low signal-to-background of three-bounce photons.
  \item Estimation of line-of-sight scene parameters.
\end{itemize}
First, only a small fraction of the emitted photons are captured in the measurement. This limits the acquisition time for NLOS imaging. Many recent works still require minutes to an hour of acquisition time for diffuse objects, which limits the real-time applications. Second, the majority of the existing works treat the line-of-sight scene as a known geometry, given that the line-of-sight scene is much easier to recover than the hidden scene. However, a fully-automated procedure to recover both line-of-sight and non-line-of-sight geometry is necessary for the practical applications of NLOS imaging, where the imaging platform could be moving.

\Paragraph{Coherence:} The coherence-based approaches exploit the observations that certain coherent properties are preserved after the scattering on the diffuse wall. The challenge towards practical speckle-based methods are:
\begin{itemize}
  \item Small field-of-view due to memory effect.
  \item Lack of depth information. 
\end{itemize}
Correlography techniques exploit memory effect, which has a limited angular field of view~\cite{Katz14}. This limits its application to the macroscopic scenes such as autonomous vehicles. Speckle-based reconstruction methods recover the 2D projection of the image but do not recover the depth.

Spatial coherence-based methods do not suffer from the small field of view as the speckle-based methods, but they require sensitive interferometric detectors that are mechanically translated~\cite{Batarseh18}. 

\Paragraph{Intensity:} The ease of image acquisition is the main advantage of intensity-based approaches. Detection of hidden moving objects around corners using shadow was demonstrated on an autonomous wheelchair~\cite{Naser2018}. The challenge towards practical intensity-base NLOS imaging are:
\begin{itemize}
  \item Unknown occluding geometries in the hidden scene.
  \item Separation of the background and signal photons.
  \item Low signal-to-noise ratio.
\end{itemize}
First, the quality of passive, occlusion-based NLOS imaging heavily depends on the occluding geometries. For example, if there are no occluding geometries in the hidden scene, the reconstruction is extremely challenging~\cite{Thrampoulidis2018ExploitingOI, Saunders2019Periscopy}. However, such information about occlusions might not be available in prior. Second, ambient illumination may be much stronger than the signal light, which makes it hard to isolate the signal from the measurement. This requires active illumination, background subtraction, or motion of the hidden scene to extract the signal. Lastly, the algorithm has to be sensitive enough to capture the small signal from the hidden scene while robust enough to reject the intensity variation due to the noise.

\subsection{Limitations on Reconstruction:} 
ToF-based NLOS imaging has a derived resolution limit. Reza et al.~\cite{reza2018physical} showed that virtual optics formulation of NLOS imaging gives the following Rayleigh lateral resolution limit:
\begin{equation}
    \Delta x \approx 1.22\frac{c\tau L}{d},
\end{equation}
where $c, \tau, L, d$ are the speed of light, full-width-half-max of the temporal response of the imaging system, the distance between the wall and hidden object, and diameter of the virtual aperture (scanning area) respectively. However, it is hard to evaluate resolution for some NLOS imaging techniques. The reconstruction capability of occlusion-based techniques heavily relies on scene geometry. It is essential to evaluate different methods with common datasets such as those proposed in ~\cite{Klein2018AQP}. 

Liu et al.~\cite{Liu_2019_CVPR} showed that the hidden object becomes impossible to reconstruct depending on the direction of its surface normal. It is also shown that the reconstruction of multiple objects may fail with a simple linear light transport model when the signal from one object is stronger than the others~\cite{Heide:2019:OcclusionNLOS}. Further theoretical investigation of such limitations is necessary for the practical use of NLOS imaging.

\subsection{Acquisition Time}
\begin{figure}
        \centering
        \includegraphics[width=0.95\linewidth]{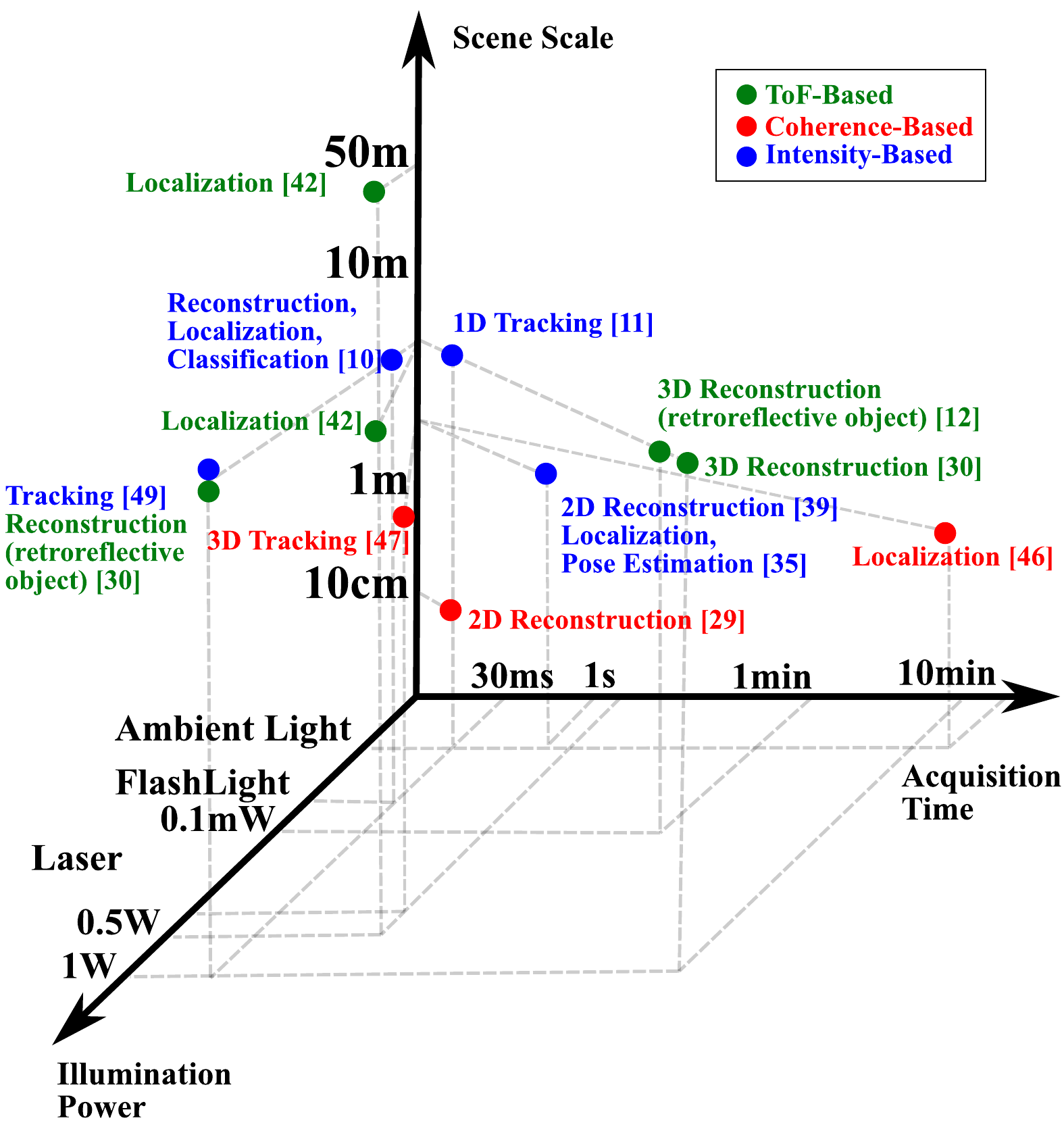}
        \caption{\textbf{Overview of the state-of-art NLOS imaging demonstration across different acquisition time, illumination power, and scene scale.}}
    \label{fig:power_time_scale.pdf}
    \vspace{-2mm}
\end{figure}

The low signal-to-noise ratio (SNR) or signal-to-background ratio is a common issue for most NLOS imaging techniques, as discussed in the previous section. This is a fundamental problem, as the number of photons from the hidden object is typically small. This limits the possible acquisition time necessary for satisfactory reconstruction and inference. Fig.~\ref{fig:power_time_scale.pdf} summarizes reported demonstration of NLOS imaging techniques across different scales of the acquisition time, illumination power, and scene.

\Paragraph{Active Sensing:} Most of the emitted photons will not be captured by the camera. Moreover, the 3-bounce photons that contain the information about the hidden scene consist only of a small fraction of the captured photons. Current demonstrations of real-time NLOS imaging with eye-safe power level is limited to retroreflective objects. Recent work on ToF-based NLOS imaging uses a laser of up to 1W power, but yet requires 10 minutes of acquisition time~\cite{Lindell:2019:Wave}. More than 10W power is necessary to perform reconstruction under a minute with the same quality, but increasing laser power is not scalable for safety and cost. One could use light at a different wavelength, such as near-infrared radiation, with higher eye safety power. Recently, a new scanning method was proposed to focus on a single voxel on the hidden scene, which can improve the SNR for a specific region of interest~\cite{Adithya2019:SNLOS}.

\Paragraph{Passive Sensing:} In many practical scenarios, most of the passively captured photons are ambient light, which does not interact with the hidden objects. This results in low SNR as well as the necessity to separate ambient photons and photons from the hidden objects. The signal photons can be isolated by exploiting the movements in the hidden scene~\cite{Bouman17}, or background subtraction when a measurement without the hidden object is available. Priors on the ambient illumination can also be exploited to remove the ambient term from the measurement~\cite{Seidel2019}. Because uncontrolled ambient light is hard to model, adding an active illumination to the passive technique may improve the signal to noise ratio~\cite{Tancik2018FlashPF,Chandran2019}.    

\subsection{Integration of Multi-Modal Techniques}
The measurements from different sensing modalities can be jointly exploited for NLOS imaging. Becks et al.~\cite{Beckus2018MultimodalNP} showed that fused intensity measurement and coherence measurement in a single optimization framework produces better reconstruction results. 

Insights from different approaches can be shared. For example, occlusions are not exploited in the ToF-based approaches. Occlusion-based techniques might be able to narrow down the region of interest in the hidden scene to make the ToF-based reconstruction faster.

\subsection{Data-driven Approach} We refer ``data-driven'' approach to the use of data to model the mapping between the measurement and the hidden scenes, and to produce priors on the solution to the inverse problem.

\paragraph{Learned Priors:}
Handcrafted priors such as total-variation are often used for NLOS reconstruction. More flexible priors or representations of the hidden objects can be learned from relevant data. Recent results on deep learning show the use of a generative model or discriminative model to enforce the learned priors in linear inverse problem~\cite{bora17a, Lunz2018}. Theoretical connections between convolutional neural networks and dictionary learning~\cite{aberdam2019mlcsc} suggest the suitability of emerging deep learning methodology to incorporate learned representations as priors to solve inverse problems. 

\paragraph{End-to-End Learning:} In recent years, deep learning has demonstrated a powerful set of techniques with applications to pattern recognition and studied for computational imaging applications such as CT and MRI~\cite{Jin17}. If there are correlations present in a dataset, then deep learning can offer powerful and flexible ways to approximate the function that maps the desired outputs from novel inputs. With a large dataset, a deep learning model can learn the mapping between a measurement and the desired inference (e.g., reconstruction and localization). The same network may also easily be adapted to learn the forward model as well by swapping the inputs and outputs and making changes to the model architecture. Potentially, a data-driven approach could offer calibration-free, efficient, and flexible solutions to NLOS imaging problems \cite{Tancik2018FlashPF, Tancik2018cosi, chen_2019_nlos}. Fig.~\ref{fig:data_driven} shows the use of deep learning for localization, identification, and reconstruction around corners.

Many NLOS imaging methods attempt to find efficient and robust algorithms to solve the optimization problem shown in Eq.~\ref{eq:setup}. In contrast, the data driven approach attempts to solve the following minimization problem to find the function $\hat{f}$, which approximates the mapping between the measurement $y$, and the target $x$ (3D shape, location, class of the hidden object):

\begin{equation}
\hat{f} = \arg \min_{f \in \mathcal{F}}\frac{1}{n} \sum_{i=1}^{n} l(f(y_i),x_i) + r(\theta), 
\label{eq:deeplearningloss}
\end{equation} 

where $f$ is parameterized by $\theta$ and the regularizer $r(\theta)$ discourages over-fitting. $\frac{1}{n} \sum_{i=1}^{n} l(f(y_i),x_i)$ denotes the error between the reconstruction and ground truth.

The main problems in the data-driven approach are generalizability and explainability.

\Paragraph{Generalizability:} The potential advantage of data-driven approaches is that they can learn models that are more robust to scene variation than brittle model-based approaches. For example, a data-driven approach may handle variations in a wall reflectance, while a model-based approach is often limited to simple, diffuse reflectance. However, the key concern is data-driven approaches fail in unpredictable ways when encountering scenes that are not represented in the training data.

One approach towards generalization is to produce a sufficiently large dataset that contains rich variations such that any possible practical scenes are represented in the training dataset. However, this is ultimately limited by the ability to simulate or collect enough experimental data to sufficiently sample the parameter space. Another approach is to design flexible neural networks for specific scene types. This makes it easier for the neural networks to learn the algorithm, which robustly works if the scene type is identified correctly.

\Paragraph{Explainability:}
Another challenge for data-driven approaches is that Eq.~\ref{eq:deeplearningloss} produces a mapping that is difficult to interpret. Rather than incorporating a known forward model, the learned mapping is based on the statistics present in the training set and is susceptible to ``hallucinating'' output in order to produce a ``reasonable output'' with respect to the training data. 

Instead of learning Eq.~\ref{eq:deeplearningloss} directly, a combination of physics-based forward models and data-driven approaches could make these methods more explainable~\cite{Che2018ioannisinversenetworks}. To solve the optimization problem in Eq.~\ref{eq:setup}, iterative algorithms such as ADMM and ISTA demonstrate improved performance when learning priors or forward models from a specific distribution~\cite{Gregor2010LearningFA, Yan16:ADMM-net, Chang17}. While traditional model-based approaches typically rely on a handcrafted sparsity prior, a data-driven approach could offer a prior that more accurately models the scene distribution~\cite{Lunz2018, aberdam2019mlcsc}. Furthermore, these algorithms can be embedded within the architecture of the deep learning model directly~\cite{Diamond2017deepprior}. Incorporating data-driven techniques into the traditional model-based optimization schemes show promise for NLOS imaging without sacrificing explainability.

\section{Conclusions and Future Directions}
We reviewed the existing NLOS imaging techniques that rely on different principles and discussed challenges towards the practical, real-time NLOS imaging. We hope that this paper will help and inspire further research toward practical NLOS imaging.

\section{Acknowledgement}
The authors thank Adithya Pediredla, Ravi Athale, and Sebastian Bauer for valuable feedback. This work is supported by DARPA REVEAL program (N0014-18-1-2894) and MIT Lab consortium funding.
\bibliographystyle{unsrt}
\bibliography{main}

\end{document}